# Challenges of Critical and Emancipatory Design Science Research
## *The Design of 'Possible Worlds' as Response*

J. Marcel Heusinger

*Institute for Computer Science and Business Information Systems, University of Duisburg-Essen,
Universitaetsstr. 9, Essen, Germany*

*marcel.heusinger@icb.uni-due.de*

Abstract: Popper's (1967) 'piecemeal social change' is an approach manifesting itself in science as critical and emancipatory (C&E) research. It is concerned with incrementally removing manifested inequalities to achieve a 'better' world. Although design science research in information systems seems to be a prime candidate for such endeavors, respective projects are clearly underrepresented. This position paper argues that this is due to the demand of justifying research ex post by an evaluation in practical settings. From the perspective of C&E research it is questionable if powerful actors grant access to their organization and support projects which ultimately challenge their position. It is suggested that theory development based on a synthesis of justificatory knowledge is a complementary approach that allows designing realizable responses to C&E issues–the design of 'possible worlds' (Lewis, 1986) as basis for C&E design science research.

## 1 INTRODUCTION

Critical and emancipatory (C&E) research projects are one of three application areas of information systems research (ISR) (Iivari, 2007). However, within ISR in general and design science research in information systems (DSRIS) in particular, there is a clear lack of such projects (Carlsson, 2010; Myers & Klein, 2011). This is puzzling because DSRIS with its aim of changing existing structures and processes (Iivari, 2007, 2010; Purao et al., 2010; Sein et al., 2007) seems to be a prime candidate for this endeavor. This is most obvious in the research stream which conceptualizes information systems (IS) as socio-technical systems (e.g., Carlsson, 2007, 2010; Carlsson et al., 2011; Hevner, 2007; Hevner, et al., 2004; Österle et al., 2010, 2011; Venable, 2006; Walls et al., 2004). This stream conceives information and communication technology (ICT) applications as an element embedded in an action system, comprising human beings and processes, and does not, as the much narrower view, exclude almost anything but the ICT application (e.g., Gregor, 2009; Kuechler & Vaishnavi, 2012a, 2012b; Nunamaker et al., 1991; Peffers et al., 2008). Although both conceptualizations inevitably transform action systems to IS or change existing IS, the broader perspective not only recognizes these changes in composition and structure, it also allows to deliberately plan them. This can, in reference to Lewis (1986), be called the design of 'possible worlds', which were introduced to ISR by Frank (2009). As the idea of a nomologically 'possible world' is a prerequisite for questioning existing structures and processes (Frank, 2009; Zelewski, 2007), it provides the basis for C&E projects concentrating upon the identification and removal of manifested injustices (Robson, 2002). In addition, DSRIS has the unique potential to form the methodological foundation for building means to overcome the identified injustices.

Correspondingly, it seems worthwhile investigating how DSRIS can be leveraged for C&E research. This position paper therefore sketches the idea of an approach focusing on the design of 'possible worlds' as a response to C&E issues. One part of the thesis advocated in the present paper is that a 'realist synthesis' (Pawson, 2006) is a theorizing technique, which allows to gather justificatory, design-relevant knowledge from practical, theorizing, and theoretical ISR as well as from relevant reference disciplines and that this body of knowledge informs the selection and development of two mid-range design theories, viz. information systems design theories (ISDTs) (Walls et al., 1992) and design-relevant explanatory and predictive theories (DREPTs) (Kuechler & Vaishnavi, 2012a). The second part of the thesis is that theorized 'possible worlds' represent a self-contained C&E research project.

Outlining the position underpinning the development of a corresponding approach is highly relevant because it reflects the methodological self-conception of DSRIS. However, relevance is a characteristic attributed by the target audience (Frank, 2006), who has to pass the final judgment. The primary audiences of this position paper are scientists, especially those who want to conduct C&E projects as well as, but to a lesser extent, those concerned with theory development in DSRIS.

The remainder is structured as follows: In the succeeding section two anticipated responses to the above-mentioned thesis are presented and discussed. Based on this preparatory work, the third section puts forward three arguments outlining the approach for designing 'possible worlds', which is currently being developed by the author. The final section concludes the discussion.

## 2 CHALLENGES OF C&E RESEARCH

DSRIS sets out as a paradigm bridging both, 'relevance' and 'rigor'. Responses to the thesis stated in the previous section evolving from those concerns seem to be the most serious. Therefore, this section deals with two respective counterclaims: (1) artifacts need to be rigorously evaluated to justify the 'effectiveness' or 'validity' of the implied claim and (2) relevant research deals with problems and opportunities articulated in practice.

The first counterclaim seems to be the most pressuring as almost all DSRIS approaches demand an evaluation (e.g., Becker, 2010; Carlsson, 2010; Hevner et al., 2004; Kuechler & Vaishnavi 2008, 2012a; Nunamaker et al., 1991; Österle et al., 2010, 2011; Peffers et al., 2008; Venable, 2006). The goal of an evaluation is to assess the efficacy or consequences of the artifact's instantiation in use (Gregor, 2009) by either employing empirical-quantitative (Iivari, 2010) or interpretive (Hevner & Chatterjee, 2010) methods. Instantiation and evaluation are mandatory activities for a valid research project (Riege et al., 2009). This is common tenor of DSRIS: from more general instructions such as Hevner et al.'s (2004) third guideline (i.e., "[t]he utility, quality, and efficacy of a design artifact must be rigorously demonstrated via well-executed evaluation methods") to the more specific demands of Kuechler and Vaishnavi (2012a) in theory development (i.e., the "[v]alidation of the artifact generates information that is used to assess the correctness of the entire reasoning /circumscription chain") (see also Niehaves, 2007; Hevner, 2007; March & Vogus, 2010; Nunamaker et al., 1991; Venable, 2006; Österle et al., 2010).

The ultimate concern is the 'effectiveness' or 'validity' of the claim(s) manifested in the artifact, that is, the evaluation is performed to justify all non-evident or unshared assumptions embodied in the artifact (Frank, 2010). In sum, the answer to how novel research results are justified, the central question of the context of justification (Ladyman, 2007), in DSRIS is verificatory, like the answer of the empirical-quantitative tradition (Zelewski, 2007). Justification through 'post-construction evaluation' is well-established, but not perfect. There is room for complementary approaches such as a 'within-construction justification'.

An argument for this pluralistic perspective of justification can be derived from difficulties associated with the conventional approach. The central challenge originates from the 'amplified contingency' (Frank, 2006) of DSRIS's unit of analysis leading to the insight that "the evaluation process in design science is task and situation specific" (March & Vogus, 2010). In other words, the evaluation of the effectiveness is spatially and temporally bound to a specific social context. This corresponds to the second moment of the scientific enterprise, the moment of 'open-systemic application of theory' (Bhaskar, 2008). In the 'moment of theory', the first moment, knowledge is gained in controlled environments (i.e., closed systems such as laboratories), which is then leveraged to measure or predict events in uncontrollable environments (i.e., open systems such as organizations). As it is impossible to control all influencing variables to isolate the effects of specific causes within open systems, observed events and their magnitude are always the result of multiple amplifying and/or curtailing influences. Because of the contingency of the context, the 'practical/technological utility' (Niiniluoto, 1993) ascertained in the evaluation in one context, does not guarantee practical utility in another. Furthermore, the suggestion to exclude trail-and-error descriptions from research reports to preserve the reader's motivation (Chmielewicz, 1994) makes it impossible to reconstruct and explain processes in open systems–a prerequisite to derive transcontextual knowledge. This in turn has the consequence that neither the possibility of transferring an artifact to another context nor the effectiveness of this transfer can be explained scientifically; they are based on experience or 'assumed rationality' (Bhaskar, 2008). Finally, focusing on 'practical utility' at the expense of the first moment's 'epistemtic utility' (Niiniluoto, 1993) inhibits eliminating hypotheses from the body of knowledge (Bunge, 1966; Chmielewicz, 1994), because the practical application of the artifact and its successful evaluation does not give an indication of the truth of the embedded theoretical propositions (Bunge, 1966). For example it might be possible that only some part of the theoretical knowledge embedded in the artifact holds in practice or the evaluation is successful despite false theoretical statements (i.e., spurious correlation). This in turn maintains the (insufficient) state of the knowledge base which forces DSRIS to "rely on intuition, experience, and trial-and-error methods" (Hevner et al., 2004) or 'assumed rationality'.

Relevance in DSRIS, as the basis from which the second counterclaim develops, is mainly concerned with the grounding of a DSRIS project's purpose in practical problems and opportunities (Hevner, 2007; Hevner et al., 2004; Österle et al., 2010; Rossi & Sein, 2003). These practice demands articulated by 'important stakeholders', predominantly managers responsible for deciding if organizational resources are committed to the construction, procurement, and usage of artifacts (Carlsson, 2007; Hevner & Chatterjee, 2010; Mertens, 2010), enter DSRIS projects in form of goals or context-specific requirements. According to the postulate of the 'absence of value judgments', which should ensure objectivity, justification has to be free from value judgments (Chmielewicz, 1994). A common interpretation of this demand is to be personally detached from values and solely focus on selecting the 'objectively' most effective means to achieve given goals. This move is possible because values have no binding force (Niiniluoto, 1993). In reference to Habermas (1987) this perspective can be called 'purposive or means-end rationality'. An extension of this type of rationality–'normative rationality'– would discuss goals and means in reference to commonly shared and acceptable social values.

Such an extended perspective seems reasonable, because science in general and applied sciences in particular have considerable societal consequences or side-effects. North (1990), awarded with the Nobel Memorial Prize in Economic Sciences in 1993, for example, argues that introducing new technology often leads to the "deliberate deskilling of the labor force", that is, highly skilled employees, with high bargain power, are substituted with less skilled and less powerful employees (for further ICT related arguments see Fountain, 2001; Stahl, 2009).

Chmielewicz (1994) argues that it is hard to accept that researchers, despite these societal consequences, work on goals and means without a normative position. He further argues that, because researchers' obligations are different from those of politicians and managers, they should consider the normative implications of their research. Similarly, Niiniluoto (1993) notes that a researcher "contributing to applied science is *morally responsible* for" his or her contribution. The exclusion of 'normative rationality', by solely focusing on 'purposive rationality' implies that human beings, an immanent part of IS, are merely treated as objects. To some degree and in special circumstances such a perspective might be acceptable for analytical purposes; however, it is a serious deficit if normative considerations are completely excluded, especially from applied disciplines. It not only makes the discipline morally questionable, it also confines intellectual curiosity—the source of important scientific problems (Bunge, 1966)—to purposive rationality; it makes demarcation of DSRIS and consulting/design practice fuzzy; and it neglects the duty of scientists to enlighten society (Albert, 1972).

This is not a call to fundamentally revise the foundations of the discipline and its methodological repertoire, but to recognize the inherent 'imperfect obligation'. To make ISR more accountable to one of its largest stakeholders, viz. society at large, the issues considered in DSRIS need to be extended. Within the next section the idea of a complementary approach focusing on the design of 'possible worlds' as solution to the identified issues is sketched.

## 3 THE DESIGN OF 'POSSIBLE WORLDS' AS RESPONSE

One implication of the previous discussion is that C&E projects in DSRIS are inevitably theorizing efforts. Therefore, it seems vital to relate the proposal to theory development in DSRIS, i.e., the framework proposed by Kuechler and Vaishnavi (2012a). In particular, two minor, closely related extensions to address the above-mentioned issues are suggested: a methodological and a conceptual. Based on these extensions, the final part of this section sketches an idea to distinguish possible and utopian worlds, to address utopism as further possible counterclaim.

Kuechler and Vaishnavi (2012a) provide a list of techniques used in theory development, which needs to be complemented by an additional research strategy that takes the peculiarities of C&E DSRIS into account: the design of 'possible worlds' requires a methodological foundation that allows justification within construction, primarily because it seems to be unlikely that those in power grant access to their organization and support a project, such as a C&E project, which ultimately challenges their position.

Within policy design and evaluation, a discipline concerned with interventions in action systems, and as such quite close to DSRIS, a successfully applied research strategy is provided by Pawson (2006). Based on earlier works (Pawson & Tilley, 1997; Tilley, 2000), he develops a 'realist synthesis', which allows gathering design knowledge for social interventions. Although this technique is mainly concerned with policy interventions, which do not necessarily entail ICT, there are no obstacles to include ICT and fruitfully apply it in DSRIS (see Carlsson, 2007, 2009, 2010; Carlsson et al., 2011). It is suggested that this technique provides the basis for a 'with-construction' justification, because it synthesizes justificatory knowledge from practical and theoretical research, which can be leveraged in the design of 'possible worlds'.

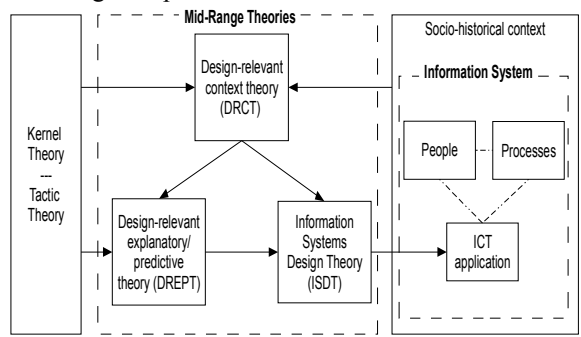

Figure 1: The Relationship between DRCT, DREPT, ISDT, and Information Systems (modified from: Kuechler and Vaishnavi (2012a)).

The discussion in the preceding section further indicates that designed ICT applications do not exist in a vacuum, but are embedded in an application domain. This suggests two extensions of Kuechler and Vaishnavi's (2012a) framework: (1) a broader view of IS, comprising people and processes in addition to ICT applications; and (2) the inclusion of the socio-historical context which can account for the 'path dependency' (David, 1985) of action and socio-technical systems. Both these extensions are depicted on the right side of figure 1. A second conceptual extension is the inclusion of design-relevant context theories (DRCTs), which capture the results of the above-mentioned synthesizing efforts. They are similar to what Kuechler and Vaishnavi (2012a) define as DREPT, however, DRCTs are not issue-centered like DREPTs, but describe the context that specify the meta-requirements in ISDTs (Walls et al., 1992) and influence the selection of DREPTs for the development of ISDTs. For example, Walls et al. (1992) derive their "how to manage", which enters into the meta-requirements, from (i) "how people should manage" and (ii) "how people manage". Whereas (i) indicates the connection between kernel theories and DRCTs, (ii) refers to the connection between IS and DRCTs. Understanding available options for intervening with an ICT application in an IS and assess the potential success requires identifying relevant high-level institutional (e.g., country-specific and international policies) and historical influences (e.g., societal norms) as both shape the effectiveness of ICT applications. Furthermore, DRCTs also capture possible challenges in "how to manage" for which DREPTs, such as the ones discussed by Kuechler and Vaishnavi (2012a), are selected to integrate features into ICT applications, which allow overcoming these issues. Hence, DRCTs not only influence the development of ISDTs, they also connect multiple appropriate DREPTs used in their development. This supports the 'artifact's mutability' (Gregor and Jones, 2007).

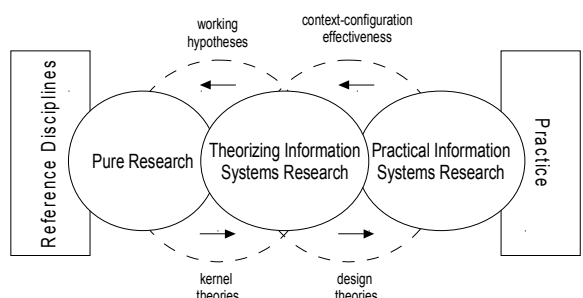

Figure 2: Three Roles of Information Systems Researchers.

These extensions provide the basis for the design of 'possible worlds'. Generally, 'possible worlds' are too complex to be achieved in a single step and (therefore) require multiple intermediate interventions, each creating a different context. The various 'context shifts', eventually culminating in the 'possible world', are captured by DRCTs. For example, an obstacle to a 'possible Open Access (OA) world' might be the concern about the review quality, which causes authors to publish their articles in closed access journals. A simple, successful intervention is disclosing the reviewers' names to provide an additional incentive. Introduced into a particular context this intervention transforms the context to one without the quality concern; however, it leaves other issues untouched (e.g., high costs of OA journals), which can be addressed by ICT-based interventions. Following Gregor (2009) and Iivari (2010) it is argued that such theorizing projects constitute self-contained research projects, even without a post-construction evaluation. Instead, each context shift is justified, as far as possible, in reference to appropriate synthesized research results. Propositions for which appropriate studies are not yet available have to be labeled as 'working hypotheses' (Frank, 2010), the subject of further pure or theoretical research (see figure 2). The results of such efforts are later (re-)integrated into DRCTs via kernel theories. Furthermore, as the synthesized justificatory knowledge focuses on lower-level propositions, the total effect, assumed to lead to the desired 'possible world', has to be tested and refined in practical research. The gained insights and context-specific adaptions or case differentiations (context-configuration effectiveness in figure 2) are the basis for further synthesizing, eventually leading to more robust and refined DRCTs.

Finally, to avoid the utopism counterclaim, a potential allegation in response to the inclusion of working hypotheses, it has to be shown that the 'possible world' is in fact possible. This requires to justify that the change leading the desired 'possible world' is potentially realizable (Frank, 2009). Following Chmielewicz (1994) this can be called the realization hurdle: is the proposed alternative realizable or possible? The synthesis needs to justify that this hurdle can be overcome by providing evidence for the following five questions (Chmielewicz, 1994): is the change (1) logically, (2) theoretically (based on natural and social laws), (3) instrumentally (technological), (4) economically, and (5) normatively possible? The amount of justificatory evidence gathered to answer these questions determines how likely it is to realize the 'possible world'. Ideally, sufficient evidence is provided for all these issues, in a 'real' theorizing project, however, the effort has to be aligned with the intention as well as the available resources, i.e., only conceptually possible 'possible worlds', as subset of all logically possible 'possible worlds', tends to be of interest to C&E DSR.

## 4 CONCLUSIONS

The main argument put forward in the position paper is that the conventional conceptualization of DSRIS, especially the demand to evaluate an instantiated artifact in a practical setting, tends to disadvantage C&E projects. The principal reason is that powerful actors will not support projects endangering their position. The accustomed response of researchers is to detach themselves from given goals and focus on the practical problems and opportunities articulated by the powerful. This neglects the (imperfect) obligation to consciously consider the social consequences of technological interventions. Although such a move might be acceptable in certain circumstances, a discipline completely excluding C&E endeavors is morally questionable. Therefore, it was suggested to extend the methodological foundation of DSRIS in such a way that these issues can be addressed. It was argued that exploring the design of 'possible worlds' is a fruitful direction for identifying complementary approaches. A brief sketch of a method based on the synthesis of justificatory knowledge from practical and theoretical research was given. Based on the quoted literature and the arguments put forward in this position paper, a detailed procedure for the design of 'possible worlds' is currently being developed by the author.